\documentclass[12pt,ams]{article}
\usepackage{amsfonts}
\usepackage{graphicx}
\usepackage{amsmath}

\begin{document}

\title{\textbf{Gribov horizon and BRST symmetry: a few remarks} }
\author{\textbf{S.P. Sorella}\thanks{
sorella@uerj.br} \thanks{%
Work supported by FAPERJ, Funda{\c c}{\~a}o de Amparo {\`a} Pesquisa do
Estado do Rio de Janeiro, under the program \textit{Cientista do Nosso Estado%
}, E-26/100.615/2007.} \\
\\
\textit{UERJ, Universidade do Estado do Rio de Janeiro,} \\
\textit{Instituto de F{\'\i}sica, } \\
\textit{\ Departamento de F{\'\i}sica Te{\'o}rica, } \\
\textit{Rua S{\~a}o Francisco Xavier 524,} \\
\textit{20550-013 Maracan{\~a}, Rio de Janeiro, Brasil}}
\maketitle

\begin{abstract}
\noindent The issue of the BRST symmetry in presence of the Gribov horizon
is addressed in Euclidean Yang-Mills theories in the Landau gauge. The
positivity of the Faddeev-Popov operator within the Gribov region enables us
to convert the soft breaking of the BRST invariance exhibited by the
Gribov-Zwanziger action into a non-local exact symmetry,
displaying explicit dependence from the non-perturbative Gribov parameter.
Despite its non-locality, this symmetry turns out to be useful in order to
establish non-perturbative Ward identities, allowing us to
evaluate the vacuum expectation value of quantities which are BRST\ exact.
These results are generalized to the refined Gribov-Zwanziger action
introduced in \cite{Dudal:2007cw}, which yields a gluon propagator
which is non-vanishing at the origin in momentum space, and a ghost propagator
which is not enhanced in the infrared.
\end{abstract}

\pagebreak

\section{Introduction}

The BRST\ symmetry is a fundamental tool for the study of gauge theories.
This symmetry is at the origin of the Slavnov-Taylor identities which
guarantee the renormalizability of Yang-Mills
theories \cite{Becchi:2005dr}. Furthermore, the use of the BRST\ charge
enables us to identify the physical subspace of the theory, allowing to
prove the unitarity of the $S-$matrix\footnote{%
Here, we are referring to theories for which the asymptotic fields can be
consistently introduced.}. Though, reconciling this
symmetry with the appearance of non-perturbative
effects represents a non-trivial challenge in quantum field theory. \newline
\newline
In particular, the study of the BRST symmetry when the Gribov
copies \cite{Gribov:1977wm} are taken into account is of great
relevance in order to unravel the non-perturbative features of
confining Yang-Mills theories \cite{Fujikawa:1982ss,Dudal:2009xh}.
While the Gribov copies can be neglected in the perturbative
ultraviolet regime, they play a relevant role in the infrared,
being closely related to gluon confinement. A partial resolution
of the Gribov issue is provided by the
Gribov-Zwanziger framework \cite%
{Gribov:1977wm,Zwanziger:1989mf,Zwanziger:1992qr}, amounting to restrict the
domain of integration in the functional integral to the Gribov region, whose
boundary is the first Gribov horizon. Although the Gribov region is not
completely free from Gribov copies \cite{vanBaal:1991zw}, the restriction to
this region can be achieved by adding to the Yang-Mills action a non-local
term $S_{h}$, known as the horizon function \cite%
{Zwanziger:1989mf,Zwanziger:1992qr}
\begin{eqnarray}
S_{h} &=&\gamma ^{4}\int d^{4}x\;h(x)\;,  \notag \\
h(x) &=&g^{2}f^{abc}A_{\mu }^{b}\left( \mathcal{M}^{-1}\right)
^{ad}f^{dec}A_{\mu }^{e}\;,  \label{m1}
\end{eqnarray}%
where $\left( \mathcal{M}^{-1}\right) ^{ad}$ is the inverse of the
Faddeev-Popov operator $\mathcal{M}^{ab}=-\partial _{\mu
}\left( \partial _{\mu }\delta ^{ab}+gf^{acb}A_{\mu }^{c}\right) $. The
massive parameter $\gamma $ appearing in eq.$\left( \ref{m1}\right) $ is the
Gribov parameter \cite{Gribov:1977wm}. It is not a free parameter, being
determined in a self-consistent way as a function of the gauge coupling
constant $g$ and of the invariant scale $\Lambda _{QCD}$ through the so
called horizon condition \cite{Zwanziger:1989mf,Zwanziger:1992qr}
\begin{equation}
\left\langle h(x)\right\rangle =4(N^{2}-1)\;.  \label{m3}
\end{equation}%
Despite its non-locality, the horizon function can be cast in
local form by means of a set of auxiliary fields, $\left(
\overline{\varphi }_{\mu }^{ab},\varphi _{\mu
}^{ab},\overline{\omega }_{\mu }^{ab},\omega _{\mu }^{ab}\right)
$, see Sect.3. A local action is thus obtained from the Gribov
horizon. Remarkably, it enjoys the important property of being
renormalizable
\cite{Zwanziger:1989mf,Zwanziger:1992qr,Maggiore:1993wq,Dudal:2005na,Dudal:2008sp,Gracey:2006dr}.\newline
\newline
Having at our disposal a local renormalizable action encoding the restriction to the Gribov
region, we can ask ourselves whether it possesses BRST\ invariance. Here, we
need to be more precise about the question we are addressing. Till now, the
BRST\ transformations we are referring to are those corresponding to the
Faddeev-Popov Lagrangian with the inclusion of the auxiliary fields needed
to localize the horizon function.  As pointed out in \cite%
{Zwanziger:1989mf,Zwanziger:1992qr}, these fields give rise to a BRST
quartet. Thus, we are considering the following BRST\ operator $s$:%
\begin{align}
sA_{\mu }^{a}& =-D_{\mu }^{ab}c^{b}\;, & sc^{a}& =\frac{1}{2}%
gf^{abc}c^{b}c^{c}\;,  \notag \\
s\overline{c}^{a}& =b^{a}\;, & sb^{a}& =0\;,  \notag \\
s\varphi _{\mu }^{ab}& =\omega _{\mu }^{ab}\;, & s\omega _{\mu }^{ab}& =0\;,
\notag \\
s\overline{\omega }_{\mu }^{ab}& =\overline{\varphi }_{\mu }^{ab}\;, & s%
\overline{\varphi }_{\mu }^{ab}& =0\;,  \label{ii1}
\end{align}%
where $b^{a}$ is the Lagrange multiplier enforcing the gauge condition and $%
\left( \overline{c}^{a},c^{a}\right) $ stand for the Faddeev-Popov
ghosts. As a first answer to the previous question, one can check
if the BRST\ transformations $\left( \ref{ii1}\right) $ leave the
Gribov-Zwanziger action invariant. This is not the case. It turns
out that the Gribov-Zwanziger action does not exhibit invariance
under the BRST\ transformations $\left( \ref{ii1}\right) $, which
are broken by the presence of the horizon\footnote{ See Chapter V
of \cite{Dudal:2008sp} for a detailed discussion on the soft
breaking of the BRST symmetry.}. Nevertheless, the resulting
breaking term displays peculiar features. It is proportional to
the Gribov parameter $\gamma $, being thus a soft breaking. As
such, it can be neglected in the deep ultraviolet region, where
the notion of exact BRST symmetry is recovered. Moreover, it can
be kept under control at the quantum level, due to the possibility
of deriving generalized softly broken Slavnov-Taylor identities
ensuring the renormalizability of the Gribov-Zwanziger action.
\newline
\newline
In \cite{Dudal:2008sp}, particular attention was devoted to the auxiliary
fields $\left( \overline{\varphi }_{\mu }^{ab},\varphi _{\mu }^{ab},%
\overline{\omega }_{\mu }^{ab},\omega _{\mu }^{ab}\right) $ which,
as much as the Faddeev-Popov ghosts $\left(
\overline{c}^{a},c^{a}\right) $, develop their own dynamics,
giving rise to non-perturbative effects. We quote, for example,
the dimension two condensate $\left\langle 0\left\vert \left(
\overline{\varphi }_{\mu }^{ab}(x)\varphi _{\mu
}^{ab}(x)-\overline{\omega }_{\mu }^{ab}(x)\omega _{\mu
}^{ab}(x)\right) \right\vert 0\right\rangle $, easily
written as the vacuum expectation value of a BRST exact quantity, \textit{%
i.e.}
\begin{equation}
\left\langle 0\left\vert \left( \overline{\varphi }_{\mu }^{ab}(x)\varphi
_{\mu }^{ab}(x)-\overline{\omega }_{\mu }^{ab}(x)\omega _{\mu
}^{ab}(x)\right) \right\vert 0\right\rangle =\left\langle 0\left\vert
s\left( \overline{\omega }_{\mu }^{ab}(x)\varphi _{\mu }^{ab}(x)\right)
\right\vert 0\right\rangle \ .  \label{i1}
\end{equation}%
Without the restriction to the Gribov region, such an expectation value
would vanish. However, in the presence of the horizon, and thus for a
non-vanishing Gribov parameter $\gamma $, it does not necessarily vanish. The
explicit one-loop evaluation of expression $\left( \ref{i1}\right) $ shows
in fact that it is proportional to the Gribov parameter \cite{Dudal:2008sp}:
\begin{equation}
\left\langle 0\left\vert \left( \overline{\varphi }_{\mu }^{ab}(x)\varphi _{\mu
}^{ab}(x)-\overline{\omega }_{\mu }^{ab}(x)\omega _{\mu }^{ab}(x)\right) \right\vert
0\right\rangle =\frac{3\sqrt{2}}{64\pi }g\sqrt{N}(N^{2}-1)\gamma ^{2}\ .
\label{i3}
\end{equation}%
That the vacuum expectation value of quantities which are BRST exact is
non-vanishing has a simple understanding, due to the presence of a boundary
in field space, namely, the Gribov horizon. We observe that the nilpotent
BRST\ operator $\left( \ref{ii1}\right) $ can be considered as an exterior
derivative along the gauge orbit. However, even if the BRST exact quantity $%
s\left( \overline{\omega }_{\mu }^{ab}\varphi _{\mu }^{ab}\right) $ can be
seen as a total derivative, its integral, \textit{i.e.} $\left\langle
0\left\vert s\left( \overline{\omega }_{\mu }^{ab}\varphi _{\mu
}^{ab}\right) \right\vert 0\right\rangle $, picks up a boundary
contribution, encoded in the explicit dependence of $\left\langle
0\left\vert \left( \overline{\varphi }_{\mu }^{ab}\varphi _{\mu }^{ab}-%
\overline{\omega }_{\mu }^{ab}\omega _{\mu }^{ab}\right) \right\vert
0\right\rangle $ from the Gribov parameter. The existence of a
soft breaking of the BRST\ symmetry plays an important role here. As noticed
in \cite{Dudal:2008sp}, this breaking ensures that the parameter $%
\gamma $ is a physical parameter of the theory, entering thus the expression
of the correlation functions of the theory. \newline
\newline
The analysis of the soft BRST breaking done in \cite{Dudal:2008sp}
has also revealed that the possibility of achieving an exact
invariance through a suitable modification of the BRST operator by
{\it local} $\gamma$-dependent terms has to be ruled out, due to
the strong constraints imposed by dimensionality of the fields,
Lorentz covariance and $SU(N)$ color structure. Nevertheless, we
cannot disregard the existence of a non-perturbative {\it
non-local} symmetry of the Gribov-Zwanziger action, which would be
compatible with the presence of the Gribov horizon. This is the
issue we aim to face in this work. More specifically, we point out
that the soft breaking of the BRST\ symmetry displayed by the
Gribov-Zwanziger action can be converted into a non-local exact
invariance. In other words, we shall be able to show that the BRST
operator $s$ and its soft breaking can be replaced by an operator
$s_{\gamma }$ which corresponds to an exact invariance of the
Gribov-Zwanziger action, while exhibiting an explicit
dependence from the Gribov parameter $\gamma $. Albeit the operator $%
s_{\gamma }$ is non-local, it can be employed to
investigate the dynamics of the auxiliary fields $\left( \overline{\varphi }%
_{\mu }^{ab},\varphi _{\mu }^{ab},\overline{\omega }_{\mu }^{ab},\omega
_{\mu }^{ab}\right) $. We observe in fact that, imposing the condition
\begin{equation}
\left\langle 0\left\vert s_{\gamma }\Theta(x) \right\vert 0\right\rangle =0\ ,
\label{i4}
\end{equation}%
where $\Theta(x) $ stands for a local field polynomial depending on the auxiliary
fields $\left( \overline{\varphi }_{\mu }^{ab},\varphi _{\mu }^{ab},%
\overline{\omega }_{\mu }^{ab},\omega _{\mu }^{ab}\right) $, will provide us
a way to evaluate the BRST\ exact quantity $\left\langle 0\left\vert s\Theta(x)
\right\vert 0\right\rangle $. For example, from the condition $\left\langle
0\left\vert s_{\gamma }\left( \overline{\omega }_{\mu }^{ab}(x)\varphi _{\mu
}^{ab}(x)\right) \right\vert 0\right\rangle =0$, we shall be able to obtain the expression
of the condensate $\left\langle 0\left\vert \left(
\overline{\varphi }_{\mu }^{ab}(x)\varphi _{\mu }^{ab}(x)-\overline{\omega }_{\mu
}^{ab}(x)\omega _{\mu }^{ab}(x)\right) \right\vert 0\right\rangle $. Of course,
the value obtained for $\left\langle 0\left\vert \left( \overline{\varphi }%
_{\mu }^{ab}\varphi _{\mu }^{ab}-\overline{\omega }_{\mu }^{ab}\omega _{\mu
}^{ab}\right) \right\vert 0\right\rangle $ from condition $\left( \ref{i4}%
\right) $ coincides with that which has been found by direct calculations
\cite{Dudal:2008sp}. \newline
\newline
We also underline that condition $\left( \ref{i4}\right) $ turns
out to be compatible with the gap equation $\left( \ref{m3}\right) $
determining the Gribov parameter $\gamma $. As we shall see, from the horizon
condition $\left( \ref{m3}\right) $ it follows that
\begin{equation}
\left\langle 0\left\vert s_{\gamma }\left( gf^{abc}A_{\mu }^{a}(x)\overline{%
\omega }_{\mu }^{bc}(x)\right) \right\vert 0\right\rangle =0\ ,  \label{m5}
\end{equation}%
which is a particular case of condition $\left( \ref{i4}\right) $. \newline
\newline
The paper is organized as follows. In Sect.2, before facing the more complex
case of the Gribov-Zwanziger action, we construct a toy model exhibiting
soft breaking of the BRST\ invariance and reproducing features of the
Gribov-Zwanziger action. This model will be employed to show how the
operator $s_{\gamma }$ can be introduced and how condition $\left( \ref{i4}%
\right) $ enables us to evaluate the vacuum expectation value of BRST\ exact
quantities. The Noether current corresponding to the exact $s_{\gamma }$%
-invariance will be derived. Sect.3 is devoted to the Gribov-Zwanziger
action. We introduce the operator $s_{\gamma }$ and use it in order to
evaluate the condensate $\left\langle 0\left\vert \left( \overline{\varphi }%
_{\mu }^{ab}\varphi _{\mu }^{ab}-\overline{\omega }_{\mu }^{ab}\omega _{\mu
}^{ab}\right) \right\vert 0\right\rangle $. The relationship between
condition $\left( \ref{i4}\right) $ and the gap equation $\left( \ref%
{m3}\right) $ is also discussed. In Sect.4 we generalize the operator $%
s_{\gamma }$ to the so called refined Gribov-Zwanziger model \cite%
{Dudal:2007cw,Dudal:2008sp}, in which the operator $\left( \overline{\varphi
}_{\mu }^{ab}\varphi _{\mu }^{ab}-\overline{\omega }_{\mu }^{ab}\omega _{\mu
}^{ab}\right) $ is taken into account from the beginning. In Sect.5 we
collect our conclusion.

\section{A toy model and its BRST\ soft breaking}

Let us start by considering a scalar field $\phi $ whose dynamics is described by the
Euclidean action
\begin{equation}
S_{\mathrm{\phi }}=\int d^{4}x\left( \frac{1}{2}\phi (-\partial ^{2})\phi \ +%
\frac{\lambda }{4!}\phi ^{4}\right) \ .  \label{c1}
\end{equation}%
We look at a mechanism allowing us to modify in a smooth way the behavior of
the correlation functions of the theory in the infrared, while leaving the
ultraviolet behavior unmodified. As
pointed out in \cite{Baulieu:2008fy}, this goal can be achieved by the
mechanism of the soft breaking of the BRST\ symmetry. To that purpose, we
introduce a BRST quartet $(\overline{\varphi },\varphi,\overline{\omega },\omega)$
\begin{eqnarray}
s\varphi &=&\omega \ ,\ \ \ \ \ \ s\omega =0\ ,  \notag \\
s\overline{\omega } &=&\overline{\varphi }\ ,\ \ \ \ \ s\overline{\varphi }%
=0\ ,  \label{a1}
\end{eqnarray}%
and
\begin{equation}
s\phi =0\ ,  \label{a2}
\end{equation}
so that the operator $s$ is nilpotent
\begin{equation}
s^{2}=0\ .  \label{a2b2}
\end{equation}%
The fields $\left( \overline{\omega },\omega \right) $ are anti-commuting and
have ghost number $(-1,1)$, while $\left( \overline{\varphi },\varphi
\right) $ are a pair of complex conjugate commuting fields, carrying vanishing ghost number.
For the BRST invariant action we write
\begin{equation}
S_{\mathrm{inv}}=S_{\mathrm{\phi }}+S_{\mathrm{exact}}\ ,
\end{equation}%
where
\begin{eqnarray}
S_{\mathrm{exact}} &=&s\int d^{4}x\left( \overline{\omega }(-\partial
^{2})\varphi +\rho ^{2}\overline{\omega }\varphi \right) \   \notag \\
&=&\int d^{4}x\left( \overline{\varphi }(-\partial ^{2})\varphi -\overline{%
\omega }(-\partial ^{2})\omega +\rho ^{2}\left( \overline{\varphi }\varphi -%
\overline{\omega }\omega \right) \right) \ .\   \label{a4}
\end{eqnarray}%
The action $S_{\mathrm{inv}}$ has the same physical content of $S_{\mathrm{%
\phi }}$. This follows by observing that the integration over the fields $%
\left( \overline{\varphi },\varphi ,\overline{\omega },\omega \right) $
amounts to introduce a unity factor in the path integral. \newline
\newline
Further, we introduce a soft breaking of the BRST invariance, through the
term
\begin{equation}
S_{\vartheta }=\frac{\vartheta ^{2}}{2}\int d^{4}x\;\phi \left( \varphi -%
\overline{\varphi }\right) \ ,  \label{a5}
\end{equation}%
where $\vartheta $ is a mass parameter, here introduced by hand, which will
enable us to modify the large distance behavior of the correlation functions
$\left\langle \phi (x_{1})....\phi (x_{n})\right\rangle $. Thus, for the
starting action $S$, we have
\begin{equation}
S=S_{\mathrm{inv}}+S_{\vartheta }\ ,  \label{a6}
\end{equation}%
and
\begin{equation}
sS=\frac{\vartheta ^{2}}{2}\int d^{4}x\phi \omega \ ,  \label{a7}
\end{equation}%
showing that the BRST symmetry is softly broken. It is interesting to
note that the fields $\left( \overline{\varphi },\varphi ,\overline{\omega }%
,\omega \right) $ can be integrated out also in the presence of the breaking
term $S_{\vartheta }$, yielding the nonlocal action
\begin{equation}
\int d^{4}x\left( \frac{1}{2}\phi \left( -\partial ^{2}+\frac{\vartheta ^{4}%
}{2}\frac{1}{\left( -\partial ^{2}+\rho ^{2}\right) }\right) \phi \ +\frac{%
\lambda }{4!}\phi ^{4}\right) \ \ .  \label{a7-1}
\end{equation}%
The fields $\left( \overline{\varphi },\varphi ,\overline{\omega },\omega
\right) $ can be seen thus as auxiliary fields needed to localize the term $%
\phi \frac{\vartheta ^{4}}{\left( -\partial ^{2}+\rho ^{2}\right) }\phi $ in
expression $\left( \ref{a7-1}\right) $. In this sense, these fields play a
role analogous to that of the auxiliary fields $\left( \overline{\varphi }%
_{\mu }^{ab},\varphi _{\mu }^{ab},\overline{\omega }_{\mu }^{ab},\omega
_{\mu }^{ab}\right) $ needed to localize the horizon function of the
Gribov-Zwanziger action. In the same way, the parameter $\vartheta $ is akin
to the Gribov parameter $\gamma $. Despite the presence of the soft
breaking, eq.$\left( \ref{a7}\right) $, the action $S$ turns out to be
renormalizable. This is due to the fact that the BRST\ quartet $\left(
\overline{\varphi },\varphi ,\overline{\omega },\omega \right) $ is coupled
in a linear way to the scalar field $\phi $, as exhibited by expression $%
\left( \ref{a5}\right) $. As a consequence, the equations of motion of the
fields $\left( \overline{\varphi },\varphi ,\overline{\omega },\omega
\right) $ acquire the meaning of linearly broken Ward identities, namely
\begin{eqnarray}
\frac{\delta S}{\delta \overline{\varphi }} &=&(-\partial ^{2}+\rho
^{2})\varphi -\frac{\vartheta ^{2}}{2}\phi \ ,  \notag \\
\frac{\delta S}{\delta \varphi } &=&(-\partial ^{2}+\rho ^{2})\overline{%
\varphi }+\frac{\vartheta ^{2}}{2}\phi \ ,  \notag \\
\frac{\delta S}{\delta \overline{\omega }} &=&-(-\partial ^{2}+\rho
^{2})\omega \ ,  \notag \\
\frac{\delta S}{\delta \omega } &=&(-\partial ^{2}+\rho ^{2})\overline{%
\omega }\ .  \label{a8}
\end{eqnarray}%
These identities imply that the most general counterterm turns out to be
independent from the fields of the BRST\ quartet. It depends only on the
scalar field $\phi $, ensuring thus that the theory is renormalizable. In
other words, the soft breaking of the BRST\ symmetry can be kept under
control at the quantum level. \newline
\newline
Let us give a look at the propagators:
\begin{eqnarray}
\left\langle \phi (k)\phi (-k)\right\rangle &=&\frac{k^{2}+\rho ^{2}}{%
k^{4}+\rho ^{2}k^{2}+\frac{\vartheta ^{4}}{2}}\ ,  \notag \\
\left\langle \phi (k)\varphi (-k)\right\rangle &=&\frac{\vartheta ^{2}}{2}%
\frac{1}{k^{4}+\rho ^{2}k^{2}+\frac{\vartheta ^{4}}{2}}\ ,  \notag \\
\left\langle \phi (k)\overline{\varphi }(-k)\right\rangle &=&-\frac{%
\vartheta ^{2}}{2}\frac{1}{k^{4}+\rho ^{2}k^{2}+\frac{\vartheta ^{4}}{2}}\ .
\notag \\
&&  \label{a9}
\end{eqnarray}%
From expressions $\left( \ref{a9}\right) $ it is apparent that the correlation
functions of the scalar field, $\left\langle \phi (x_{1})....\phi
(x_{n})\right\rangle $, are modified in the infrared region, due to the
presence of the soft parameters $\left( \vartheta ,\rho \right) $. Moreover,
one also observes that, in the absence of the BRST\ breaking, \textit{i.e.}
when $\vartheta =0$, the propagator of the scalar field $\phi $ turns out to
be independent from the parameter $\rho $, which becomes an unphysical gauge
parameter. As a consequence, in the absence of the breaking, the Green's
functions $\left\langle \phi (x_{1})....\phi (x_{n})\right\rangle $ will be
independent from $\rho $ as well. However, \ for non-vanishing $\vartheta $,
the parameter $\rho $ is no longer a gauge parameter, entering the
expression of $\left\langle \phi (x_{1})....\phi (x_{n})\right\rangle$.
\newline
\newline
The framework outlined here gives us a way to introduce
infrared effects in a local and renormalizable fashion. Nevertheless, the
presence of the soft breaking of the BRST\ symmetry does not enable us to
make use of the BRST operator $s$ in order to characterize the vacuum of the
theory. Due to the lack of a conserved charge, one cannot impose that the
vacuum is annihilated by the BRST\ operator. This means that the vacuum
expectation value of quantities which are BRST\ exact does not necessarily
vanish, \textit{i.e.}
\begin{equation}
\left\langle 0\left\vert s\Theta(x) \right\vert 0\right\rangle \neq 0\ .
\label{a10}
\end{equation}%
So far, the only way to access correlation functions of the type of $\left( \ref{a10}%
\right) $ is through explicit computations. Let us give an example by
considering the dimension two condensate $\left\langle \left( \overline{%
\varphi }(x)\varphi (x)-\overline{\omega }(x)\omega (x)\right) \right\rangle
$, which has precisely the form of $\left( \ref{a10}\right) $, \textit{i.e.}
\begin{equation}
\left\langle \left( \overline{\varphi }(x)\varphi(x) -\overline{\omega }(x)\omega(x)
\right) \right\rangle =\left\langle s\left( \overline{\omega }(x)\varphi(x)
\right) \right\rangle \ .  \label{a11}
\end{equation}%
This condensate can be seen as the analogue of the condensate $\left\langle
0\left\vert \left( \overline{\varphi }_{\mu }^{ab}\varphi _{\mu }^{ab}-%
\overline{\omega }_{\mu }^{ab}\omega _{\mu }^{ab}\right) \right\vert
0\right\rangle $ occurring in the case of the Gribov-Zwanziger theory.
Expression $\left( \ref{a11}\right) $ can be obtained by differentiating the
vacuum energy $E_{v}$ with respect to the parameter $\rho $. In fact, from
\begin{equation}
e^{-VE_{v}}=\int \left[ D\Phi \right] \ e^{-S}\ ,  \label{a12}
\end{equation}%
it follows that
\begin{equation}
\frac{\partial E_{v}}{\partial \rho ^{2}}=\frac{\int \left[ D\Phi \right]
\left( \overline{\varphi }\varphi -\overline{\omega }\omega \right) \ e^{-S}%
}{\int \left[ D\Phi \right] \ e^{-S}}=\left\langle \left( \overline{\varphi }%
\varphi -\overline{\omega }\omega \right) \right\rangle \ .  \label{a13}
\end{equation}%
Making use of the dimensional regularization, the explicit evaluation of the
one-loop vacuum energy gives
\begin{equation}
E_{v}=\frac{1}{2}\int \frac{d^{d}p}{\left( 2\pi \right) ^{d}}\left( \log
\left( p^{4}+p^{2}\rho ^{2}+\frac{\vartheta ^{4}}{2}\right) -\log \left(
p^{2}+\rho ^{2}\right) \right) \ .  \label{a14}
\end{equation}%
Thus,
\begin{equation}
\left\langle \left( \overline{\varphi }\varphi -\overline{\omega }\omega
\right) \right\rangle =\left\langle s\left( \overline{\omega }\varphi
\right) \right\rangle =-\frac{\vartheta ^{4}}{4}\int \frac{d^{4}p}{\left(
2\pi \right) ^{4}}\frac{1}{\left( p^{4}+p^{2}\rho ^{2}+\frac{\vartheta ^{4}}{%
2}\right) \left( p^{2}+\rho ^{2}\right) }\ ,  \label{a15}
\end{equation}%
where the integral in the right hand side of eq.$\left( \ref{a15}\right) $
is convergent. Therefore, we see that the condensate $\left\langle
\left( \overline{\varphi }\varphi -\overline{\omega }\omega \right)
\right\rangle $ is non-vanishing for non-vanishing $\vartheta $. This example
illustrates in a simple way that the we cannot access the vacuum state of
the theory with the operator $s$. \\\\From expression $\left( \ref{a15}\right) $ one also
sees that the condensate
$\left\langle
\left( \overline{\varphi }\varphi -\overline{\omega }\omega \right)
\right\rangle $  is non-vanishing when the parameter $\rho$ is set to zero, {\it i.e.}
\begin{equation}
\left\langle \left( \overline{\varphi }\varphi -\overline{\omega }\omega
\right) \right\rangle_{\rho=0} =-\frac{\vartheta ^{4}}{4}\int \frac{d^{4}p}{\left(
2\pi \right) ^{4}}\frac{1}{\left( p^{4}+\frac{\vartheta ^{4}}{%
2}\right)p^{2} } \neq 0 \ .  \label{a16}
\end{equation}%
Even if the operator $\left( \overline{\varphi }\varphi -\overline{\omega }\omega \right) $ were not
included in the starting action, it would have been generated dynamically, as it is apparent
from eq.$\left( \ref{a16}\right) $. Therefore, the introduction of the parameter $\rho$ enables
us to take into account
the operator $\left( \overline{\varphi }\varphi -\overline{\omega }\omega \right) $ from the beginning.
As we shall see, a similar situation will be encountered in the case of the Gribov-Zwanziger action.

\subsection{The modified BRST operator}

Let us look now at the possibility of finding a modified BRST operator which
corresponds to an exact invariance of the action $S$. We shall rely on the
property that the breaking term, eq.$\left( \ref{a7}\right) $, is quadratic in
the fields, due to the fact that the BRST quartet couples linearly to the
scalar field $\phi $, see expression $\left( \ref{a5}\right) $. Let us
consider the equation of motion of the field $\overline{\omega }$, \textit{%
i.e.}
\begin{equation}
\frac{\delta S}{\delta \overline{\omega }(x)}=-(-\partial ^{2}+\rho
^{2})\omega (x)\ .  \label{aaa}
\end{equation}%
We observe that the operator $(-\partial ^{2}+\rho ^{2})$ is positive
definite, so that its inverse does exist. As a consequence, eq.$\left( \ref%
{aaa}\right) $ can be solved for $\omega (x)$, yielding
\begin{equation}
\omega (x)=-\frac{1}{(-\partial ^{2}+\rho ^{2})}\frac{\delta S}{\delta
\overline{\omega }}\equiv -\int d^{4}y\frac{1}{(-\partial ^{2}+\rho
^{2})_{xy}}\frac{\delta S}{\delta \overline{\omega }(y)}\ ,  \label{a17}
\end{equation}%
where
\begin{equation}
\frac{1}{(-\partial ^{2}+\rho ^{2})_{xy}}=\int \frac{d^{4}p}{\left( 2\pi
\right) ^{4}}\frac{1}{p^{2}+\rho ^{2}}e^{-ip(x-y)}\ .  \label{a18}
\end{equation}%
Therefore, equation $\left( \ref{a7}\right) $ can be rewritten as
\begin{equation}
sS=-\frac{\vartheta ^{2}}{2}\int d^{4}x\phi \frac{1}{(-\partial ^{2}+\rho
^{2})}\frac{\delta S}{\delta \overline{\omega }}\ ,  \label{a19}
\end{equation}%
showing that the BRST\ breaking of eq.$\left( \ref{a7}\right) $ can be cast
in the form of a contact term, \textit{i.e.} of a term related to the
equations of motion. Equation $\left( \ref{a19}\right) $ implies that the
action $S$ is left invariant by the modified nilpotent operator $s_{\vartheta }$
\begin{eqnarray}
s_{\vartheta }\overline{\omega } &=&\overline{\varphi }+\frac{\vartheta ^{2}%
}{2}\frac{1}{(-\partial ^{2}+\rho ^{2})}\phi \ , \notag \\
\ \ \ \ \ \ s_{\vartheta }\overline{\varphi } &=&0\ ,  \notag \\
s_{\vartheta }\varphi &=&\omega \ ,\ \ \ \ \ \   \notag \\
s_{\vartheta }\omega &=&0\ ,  \notag \\
s_{\vartheta }\phi &=&0\ ,  \label{a20}
\end{eqnarray}
so that
\begin{equation}
s_{\vartheta }S=0\ .  \label{a21}
\end{equation}%
The operator $s_{\vartheta }$ exhibits explicit dependence from
the
parameter $\vartheta $. Moreover, it reduces to the operator $s$ when $%
\vartheta =0$. One should notice, however, that the operator $s_{\vartheta }$
is non-local. As such, it cannot be used to analyse the renormalizability
properties of the model. Nevertheless, its existence is helpful in order to
evaluate the expectation value of BRST\ exact quantities, by requiring that
\begin{equation}
\left\langle 0\left\vert s_{\vartheta }\Theta(x) \right\vert 0\right\rangle =0\
.  \label{a23}
\end{equation}%
To give an example of the usefulness of eq.$\left( \ref{a23}\right) $, let
us compute again the condensate $\left\langle \left( \overline{\varphi }%
\varphi -\overline{\omega }\omega \right) \right\rangle $, by making use of
condition $\left( \ref{a23}\right) $. From the equation
\begin{equation}
\left\langle s_{\vartheta }\left( \overline{\omega }\varphi \right)
\right\rangle =0\ ,  \label{a24}
\end{equation}%
it follows that
\begin{equation}
\left\langle \left( \overline{\varphi }\varphi -\overline{\omega }\omega
\right) \right\rangle +\frac{\vartheta ^{2}}{2}\left\langle \phi \frac{1}{%
(-\partial ^{2}+\rho ^{2})}\varphi \right\rangle =0\ .  \label{a25}
\end{equation}%
Thus
\begin{eqnarray}
\left\langle s\left( \overline{\omega }\varphi \right) \right\rangle
&=&\left\langle \left( \overline{\varphi }\varphi -\overline{\omega }\omega
\right) \right\rangle =-\frac{\vartheta ^{2}}{2}\left\langle \phi \frac{1}{%
(-\partial ^{2}+\rho ^{2})}\varphi \right\rangle  \notag \\
&=&-\frac{\vartheta ^{2}}{2}\int d^{4}y\frac{1}{(-\partial ^{2}+\rho
^{2})_{xy}}\int \frac{d^{4}q}{\left( 2\pi \right) ^{4}}\left\langle \phi
(q)\varphi (-q)\right\rangle e^{-iq(x-y)}  \notag \\
&=&-\frac{\vartheta ^{2}}{2}\int d^{4}y\frac{d^{4}q}{\left( 2\pi \right) ^{4}%
}\frac{d^{4}p}{\left( 2\pi \right) ^{4}}\frac{1}{p^{2}+\rho ^{2}}%
\left\langle \phi (q)\varphi (-q)\right\rangle e^{-iq(x-y)}e^{-ip(x-y)}
\notag \\
&=&-\frac{\vartheta ^{2}}{2}\int \frac{d^{4}q}{\left( 2\pi \right) ^{4}}%
\frac{1}{q^{2}+\rho ^{2}}\left\langle \phi (q)\varphi (-q)\right\rangle \ .
\label{a26}
\end{eqnarray}%
Finally, from equations $\left( \ref{a9}\right) $ we obtain
\begin{equation}
\left\langle s\left( \overline{\omega }\varphi \right) \right\rangle
=\left\langle \left( \overline{\varphi }\varphi -\overline{\omega }\omega
\right) \right\rangle =-\frac{\vartheta ^{4}}{4}\int \frac{d^{4}p}{\left(
2\pi \right) ^{4}}\frac{1}{\left( p^{4}+p^{2}\rho ^{2}+\frac{\vartheta ^{4}}{%
2}\right) \left( p^{2}+\rho ^{2}\right) }\ ,  \label{a27}
\end{equation}%
in agreement with expression $\left( \ref{a15}\right) $. We see thus
that condition $\left( \ref{a23}\right) $ gives us a practical way of
evaluating correlators of the kind $\left\langle 0\left\vert s\Theta(x)
\right\vert 0\right\rangle$.

\subsection{The Noether current}

Let us conclude the analysis of the toy model by deriving the Noether
current corresponding to the exact invariance $\left( \ref{a21}\right) $. To
that purpose, we make use of the functional form of the operator $%
s_{\vartheta }$, \textit{i.e.}
\begin{eqnarray}
s_{\vartheta } &=&\int d^{4}x\ w(x)\ ,  \notag \\
w(x) &=&\left( \overline{\varphi }(x)+\frac{\vartheta ^{2}}{2}\Lambda
(x)\right) \frac{\delta }{\delta \overline{\omega }(x)}+\omega (x)\frac{%
\delta }{\delta \varphi (x)}\ ,  \label{a228}
\end{eqnarray}%
where $\Lambda (x)$ stands for
\begin{equation}
\Lambda (x)=\int d^{4}y\frac{1}{(-\partial ^{2}+\rho ^{2})_{xy}}\phi
(y)=\int d^{4}y\frac{d^{4}p}{\left( 2\pi \right) ^{4}}\frac{1}{p^{2}+\rho
^{2}}e^{-ip(x-y)}\phi (y)\ ,  \label{a29}
\end{equation}%
so that
\begin{equation}
(-\partial _{x}^{2}+\rho ^{2})\Lambda (x)=\phi (x)\ .  \label{a30}
\end{equation}%
Acting with the operator $w(x)$ on the action $S$ one finds%
\begin{eqnarray}
w(x)S &=&\left( \overline{\varphi }(x)+\frac{\vartheta ^{2}}{2}\Lambda
(x)\right) (\partial ^{2}\omega -\rho ^{2}\omega )+\left( -\partial ^{2}%
\overline{\varphi }+\rho ^{2}\overline{\varphi }\right) \omega +\frac{%
\vartheta ^{2}}{2}\phi \omega \   \notag  \label{a31} \\
&=&\overline{\varphi }\partial ^{2}\omega -\left( \partial ^{2}\overline{%
\varphi }\right) \omega +\frac{\vartheta ^{2}}{2}\left( \Lambda \partial
^{2}\omega -\rho ^{2}\Lambda \omega \right) +\frac{\vartheta ^{2}}{2}\phi
\omega  \notag \\
&=&\partial _{\mu }\left( \overline{\varphi }\partial _{\mu }\omega -\left(
\partial _{\mu }\overline{\varphi }\right) \omega \right) +\frac{\vartheta
^{2}}{2}\left( \partial _{\mu }\left( \Lambda \partial _{\mu }\omega -\left(
\partial _{\mu }\Lambda \right) \omega \right) +\left( \left( \partial
^{2}-\rho ^{2}\right) \Lambda \right) \omega \right) +\frac{\vartheta ^{2}}{2%
}\phi \omega  \notag \\
&&
\end{eqnarray}%
Making use of eq.$\left( \ref{a30}\right) $, it follows
\begin{equation}
w(x)S=\partial _{\mu }j_{\mu }\ ,  \label{a32}
\end{equation}%
where the Noether current $j_{\mu }$ is given by
\begin{equation}
j_{\mu }\ =\left( \overline{\varphi }\partial _{\mu }\omega -\left( \partial
_{\mu }\overline{\varphi }\right) \omega \right) +\frac{\vartheta ^{2}}{2}%
\left( \Lambda \partial _{\mu }\omega -\left( \partial _{\mu }\Lambda
\right) \omega \right) \ .  \label{a33}
\end{equation}

\subsection{Preliminary considerations}

Several remarks can be made from the example considered here.

\begin{itemize}
\item The mechanism of the soft breaking of the BRST\ enables us to modify
in a smooth way the correlation functions of the theory in the infrared, due
to the presence of the soft parameters $\left( \vartheta ,\rho \right) $.

\item The introduction of the BRST\ quartet and of the corresponding
breaking term are done in a way which preserves locality and
renormalizability.

\item However, due to the presence of the breaking, the BRST$\ $operator $s$
does not allow us to characterize the vacuum of the theory. Nevertheless, it
turns out that the action $S$ is left invariant by a modified\ operator $%
s_{\vartheta },$ which exhibits explicit dependence from the parameter $%
\vartheta $, while reducing to the BRST\ operator $s$ for vanishing $%
\vartheta $. Although non-local, the new operator $s_{\vartheta }$ can be
employed to obtain the explicit expression of quantities which
are BRST exact.
\end{itemize}

\section{The Gribov-Zwanziger action}

\subsection{The non-local horizon function and its localization}

Let us start by giving a short overview of the Gribov-Zwanziger framework,
enabling us to implement the restriction in the Euclidean functional
integral to the Gribov region $\Omega $ \cite{Gribov:1977wm}, defined as the
set of field configurations fulfilling the Landau gauge condition and for
which the Faddeev-Popov operator,
\begin{equation}
\mathcal{M}^{ab}=-\partial _{\mu }\left( \partial _{\mu }\delta
^{ab}+gf^{acb}A_{\mu }^{c}\right) \;,  \label{sc2-1}
\end{equation}%
is strictly positive, namely
\begin{equation}
\Omega \equiv \{A_{\mu }^{a},\;\partial _{\mu }A_{\mu }^{a}=0,\;\mathcal{M}%
^{ab}>0\}\;.  \label{sc2-2}
\end{equation}%
This amounts to adding
to the Yang-Mills action
\begin{equation}
S_{\mathrm{YM}}=\frac{1}{4}\int d^{4}xF_{\mu \nu }^{a}F_{\mu \nu }^{a}\;,
\label{sc2-3}
\end{equation}%
the horizon term \cite{Zwanziger:1989mf,Zwanziger:1992qr}
\begin{eqnarray}
S_{h} &=&\gamma ^{4}\int d^{4}x\;h(x)\;,  \notag \\
h(x) &=&g^{2}f^{abc}A_{\mu }^{b}\left( \mathcal{M}^{-1}\right)
^{ad}f^{dec}A_{\mu }^{e}\;,  \label{sc2-4}
\end{eqnarray}%
where the non-local kernel $\left( \mathcal{M}^{-1}\right) ^{ad}$ stands for
the inverse of the Faddeev-Popov operator
\begin{equation}
\mathcal{M}^{ab}(x)\left[ \left( \mathcal{M}^{-1}(x,y)\right) ^{bc}\right]
=\delta ^{4}(x-y)\delta^{ac} \ .  \label{sc2-4b}
\end{equation}%
Thus, for the partition function implementing the restriction to $\Omega $
one has \cite{Zwanziger:1989mf,Zwanziger:1992qr}
\begin{equation}
\int_{\Omega }DA\ \delta (\partial A)\ \det \mathcal{M\ }e^{-S_{YM}}=\int
DA\ \delta (\partial A)\ \det \mathcal{M\ }e^{-\left( S_{YM}+S_{h}\right) }
\;, \label{sc2-5}
\end{equation}%
where the Gribov parameter $\gamma $ is determined by the horizon condition
\cite{Zwanziger:1989mf,Zwanziger:1992qr}
\begin{equation}
\left\langle h(x)\right\rangle =4(N^{2}-1)\;.  \label{sc2-6}
\end{equation}%
As already mentioned, the non-local horizon function can be localized through the introduction of a
suitable set of additional fields \cite{Zwanziger:1989mf,Zwanziger:1992qr}
\begin{equation}
e^{-S_{h}}=\int D\varphi D{\overline{\varphi }D}\omega D{\overline{\omega }\
e}^{-S_{\mathrm{loc}}} \;,  \label{sc2-7}
\end{equation}%
with
\begin{equation}
S_{\mathrm{loc}}=\int d^{4}x\left( -\overline{\varphi }_{\mu }^{ac}\
\mathcal{M}^{ab}\varphi _{\mu }^{bc}+\overline{\omega }_{\mu }^{ac}\mathcal{M%
}^{ab}\omega _{\mu }^{bc}\right) -\gamma ^{2}g\int {d^{{4}}}x\left(
f^{abc}A_{\mu }^{a}\varphi _{\mu }^{bc}+f^{abc}A_{\mu }^{a}\overline{\varphi
}_{\mu }^{bc}\right) \;.  \label{sc2-8}
\end{equation}%
The fields $\left( \overline{\varphi }_{\mu }^{ac},\varphi _{\mu
}^{ac}\right) $ are a pair of complex conjugate bosonic fields, while $%
\left( \overline{\omega }_{\mu }^{ac},\omega _{\mu }^{ac}\right) $ are
anticommuting fields. Each of these fields has $4\left( N^{2}-1\right) ^{2}$
components. Therefore, a local action is obtained from the restriction to
the Gribov region, namely
\begin{equation}
\int_{\Omega }DA\ \delta (\partial A)\ \det \mathcal{M\ }e^{-S_{YM}}=\int
DA\ D\varphi D{\overline{\varphi }D}\omega D{\overline{\omega }~DbDcD}%
\overline{{c}}\mathcal{\ }e^{-S_{\mathrm{GZ}}}\ ,
\label{sc2-9}
\end{equation}%
where $S_{\mathrm{GZ}}$ is the Gribov-Zwanziger action, given by
\begin{eqnarray}
S_{\mathrm{GZ}} &=&S_{\mathrm{YM}}+\int d^{4}x\left( b^{a}\partial _{\mu
}A_{\mu }^{a}-\overline{c}^{a}\mathcal{M}^{ab}c^{b}\right) \;+\int
d^{4}x\left( -\overline{\varphi }_{\mu }^{ac}\ \mathcal{M}^{ab}\varphi _{\mu
}^{bc}+\overline{\omega }_{\mu }^{ac}\mathcal{M}^{ab}\omega _{\mu
}^{bc}\right)  \notag \\
&&-\gamma ^{2}g\int d{^{{4}}}x\left( f^{abc}A_{\mu }^{a}\varphi _{\mu
}^{bc}+f^{abc}A_{\mu }^{a}\overline{\varphi }_{\mu }^{bc}+\frac{4}{g}\left(
N^{2}-1\right) \gamma ^{2}\right) \;.  \label{sc2-10}
\end{eqnarray}%
The constant term, $4\left( N^{2}-1\right) \gamma ^{4}$, in expression $%
\left( \ref{sc2-10}\right) $ is introduced in order to rewrite the horizon
condition $\left( \ref{sc2-6}\right) $ as
\begin{equation}
\frac{\partial \Gamma }{\partial \gamma ^{2}}=0\ ,  \label{sc2-11}
\end{equation}%
with $\Gamma $ being the effective action obtained from $S_{\mathrm{GZ}}$%
\begin{equation}
\mathrm{e}^{-\Gamma }=\int \left[ D\Phi \right] \mathrm{e}^{-S_{\mathrm{GZ}%
}}\;.  \label{sc2-11b}
\end{equation}%
To the first order, the gap equation $\left( \ref{sc2-11}\right) $ takes the
form
\begin{equation}
1=\frac{3}{4}Ng^{2}\int \frac{d^{4}p}{\left( 2\pi \right) ^{4}}\frac{1}{%
p^{4}+2g^{2}N\gamma ^{4}}\ .  \label{sc2-111}
\end{equation}%
Let us also observe that condition $\left( \ref{sc2-11}\right) $ is
equivalent to
\begin{equation}
\left\langle 0\left\vert gf^{abc}A_{\mu }^{a}\varphi _{\mu }^{bc}\right\vert
0\right\rangle +\left\langle 0\left\vert gf^{abc}A_{\mu }^{a}\overline{%
\varphi }_{\mu }^{bc}\right\vert 0\right\rangle =-8\left( N^{2}-1\right)
\gamma ^{2}\ ,  \label{sc2-11c}
\end{equation}
which, owing to the discrete symmetry of the action $S_{\mathrm{GZ}}$
\begin{eqnarray}
\overline{\varphi }_{\mu}^{bc} & \rightarrow & \varphi _{\mu }^{bc} \; , \nonumber \\
\varphi _{\mu }^{bc} &\rightarrow & \overline{\varphi }_{\mu }^{bc} \; , \nonumber \\
b^a  & \rightarrow & (b^a-gf^{amn}\; \overline{\varphi }_{\mu }^{mc}\; \varphi _{\mu }^{nc})  \; , \label{dis}
\end{eqnarray}
becomes
\begin{equation}
\left\langle 0\left\vert gf^{abc}A_{\mu }^{a}\varphi _{\mu }^{bc}\right\vert
0\right\rangle =\left\langle 0\left\vert gf^{abc}A_{\mu }^{a}\overline{%
\varphi }_{\mu }^{bc}\right\vert 0\right\rangle =-4\left( N^{2}-1\right)
\gamma ^{2}=-\gamma ^{2}\left\langle h(x)\right\rangle \ .  \label{inc1}
\end{equation}%
Equation $\left( \ref{inc1}\right) $ shows the relevance of the dimension
two condensates $\left\langle 0\left\vert gf^{abc}A_{\mu }^{a}\varphi _{\mu
}^{bc}\right\vert 0\right\rangle $, $\left\langle 0\left\vert gf^{abc}A_{\mu
}^{a}\overline{\varphi }_{\mu }^{bc}\right\vert 0\right\rangle $, within the Gribov-Zwanziger approach.

\subsection{Soft breaking of the BRST\ invariance}

As already underlined, the auxiliary fields $\left( \overline{\varphi }_{\mu
}^{ab},\varphi _{\mu }^{ab},\overline{\omega }_{\mu }^{ab},\omega _{\mu
}^{ab}\right) $ needed to localize the horizon function $\left( \ref{sc2-4}%
\right) $ give rise to a BRST\ quartet, meaning that they are assembled in
two BRST\ doublets \cite{Zwanziger:1989mf,Zwanziger:1992qr}, see eqs.$\left( %
\ref{ii1}\right) $. Acting thus with the BRST\ operator $s$ on the action $%
\left( \ref{sc2-10}\right) $, one finds
\begin{eqnarray}
sS_{\mathrm{GZ}} &=&\int d^{4}x\ gf^{abm}\left( D_{\nu }^{bp}c^{p}\right)
\left( \left( \partial _{\nu }\overline{\varphi }_{\mu }^{ac}\right) \varphi
_{\mu }^{mc}-\left( \partial _{\nu }\overline{\omega }_{\mu }^{ac}\right)
\omega _{\mu }^{mc}\right)  \notag \\
&&-\gamma ^{2}g\int d^{4}x\ \left( c^{m}D_{\mu }^{ma}\left( f^{abc}\left(
\varphi _{\mu }^{bc}+\overline{\varphi }_{\mu }^{bc}\right) \right)
+f^{abc}A_{\mu }^{a}\omega _{\mu }^{bc}\right) \ ,  \label{sc2-13}
\end{eqnarray}%
which shows that the BRST\ symmetry is broken. One notes that the first term
in the right-hand side of expression $\left( \ref{sc2-13}\right) $ can be
written as the BRST\ variation of $gf^{abm}\left( D_{\nu }^{bp}c^{p}\right)
\left( \partial _{\nu }\overline{\omega }_{\mu }^{ac}\right) \varphi _{\mu
}^{mc}$, \textit{i.e.}
\begin{equation}
gf^{abm}\left( D_{\nu }^{bp}c^{p}\right) \left( \left( \partial _{\nu }%
\overline{\varphi }_{\mu }^{ac}\right) \varphi _{\mu }^{mc}-\left( \partial
_{\nu }\overline{\omega }_{\mu }^{ac}\right) \omega _{\mu }^{mc}\right)
=-s\left( gf^{abm}\left( D_{\nu }^{bp}c^{p}\right) \left( \partial _{\nu }%
\overline{\omega }_{\mu }^{ac}\right) \varphi _{\mu }^{mc}\right) \ , \label{tr}
\end{equation}
so that it does not correspond to a real breaking. The terms which give rise to the
true BRST\ soft breaking are those proportional to the Gribov parameter $\gamma $.
The expression $gf^{abm}\left(
D_{\nu }^{bp}c^{p}\right) \left( \partial _{\nu }\overline{\omega }_{\mu
}^{ac}\right) \varphi _{\mu }^{mc}$ \ is usually introduced in the starting
Gribov-Zwanziger action. It can be obtained through a change
of variables \cite{Zwanziger:1989mf,Zwanziger:1992qr} in expression $\left( %
\ref{sc2-9}\right) $, the corresponding Jacobian being field independent.
However, for the purposes of the present work, we keep the term
$ gf^{abm}\left( D_{\nu }^{bp}c^{p}\right) \left( \left( \partial _{\nu }%
\overline{\varphi }_{\mu }^{ac}\right) \varphi _{\mu }^{mc}-\left( \partial
_{\nu }\overline{\omega }_{\mu }^{ac}\right) \omega _{\mu }^{mc}\right) $ as it stands.
\subsection{The operator $s_{\protect\gamma }$}

To construct the generalized operator $s_{\gamma }$, we proceed as in the
toy model considered in the previous section. From the Gribov-Zwanziger
action, we get
\begin{eqnarray}
\frac{\delta S_{\mathrm{GZ}}}{\delta \overline{c}^{a}} &=&-\mathcal{M}%
^{ab}c^{b}\ ,  \notag \\
\frac{\delta S_{\mathrm{GZ}}}{\delta \overline{\omega }_{\mu }^{ac}} &=&%
\mathcal{M}^{ab}\omega _{\mu }^{bc}\mathcal{\ }\mathrm{.}  \label{sc2-14}
\end{eqnarray}%
We observe now that, within the Gribov region $\Omega $, eq.$\left( \ref%
{sc2-2}\right) $, the Faddeev-Popov operator $\mathcal{M}^{ab}$ is strictly
positive, so that its inverse $\left( \mathcal{M}^{-1}\right) ^{ab}$ does
exist. This property enables us to solve equations $\left( \ref{sc2-14}%
\right) $ for the fields $c^{b}$ and $\omega _{\mu }^{bc}$, obtaining
\begin{eqnarray}
c^{b}(x) &=&-\left( \mathcal{M}^{-1}\right) _{xy}^{bc}\frac{\delta S_{%
\mathrm{GZ}}}{\delta \overline{c}_{y}^{c}}=-\int d^{4}y\ \left[ \left(
\mathcal{M}^{-1}(x,y)\right) ^{bc}\right] \frac{\delta S_{\mathrm{GZ}}}{%
\delta \overline{c}^{c}(y)}\ ,  \notag \\
\omega _{\mu }^{dc}(x) &=&\left( \mathcal{M}^{-1}\right) _{xy}^{da}\frac{%
\delta S_{\mathrm{GZ}}}{\delta \overline{\omega }_{y\mu }^{ac}}=\int d^{4}y\ %
\left[ \left( \mathcal{M}^{-1}(x,y)\right) ^{da}\right] \frac{\delta S_{%
\mathrm{GZ}}}{\delta \overline{\omega }^{ac}(y)}\ .  \label{sc2-15}
\end{eqnarray}%
Thus, equation $\left( \ref{sc2-13}\right) $ can be rewritten as
\begin{equation}
sS_{\mathrm{GZ}}=\int d^{4}x\left( -\left( D_{\nu }^{ma}\Lambda _{\nu
}^{a}\right) _{x}\left( \mathcal{M}^{-1}\right) _{xy}^{mc}\frac{\delta S_{%
\mathrm{GZ}}}{\delta \overline{c}_{y}^{c}}-\gamma ^{2}gf^{abc}A_{\mu
}^{a}(x)\left( \mathcal{M}^{-1}\right) _{xy}^{bm}\frac{\delta S_{\mathrm{GZ}}%
}{\delta \overline{\omega }_{y\mu }^{mc}}\right) \ ,  \label{sc2-16}
\end{equation}%
where $\Lambda _{\nu }^{a}$ stands for%
\begin{equation}
\Lambda _{\nu }^{a}=-\gamma ^{2}gf^{abc}\left( \varphi _{\nu }^{bc}+%
\overline{\varphi }_{\nu }^{bc}\right) -gf^{bap}\left( \left( \partial _{\nu
}\overline{\varphi }_{\mu }^{bc}\right) \varphi _{\mu }^{pc}-\left( \partial
_{\nu }\overline{\omega }_{\mu }^{bc}\right) \omega _{\mu }^{pc}\right) \ .
\label{sc2-17}
\end{equation}%
As in the case of the toy model, we see that the soft breaking of the BRST\
symmetry turns out to be related to contact terms, enabling us to introduce
the extended operator we are looking for. Eq.$\left( \ref{sc2-16}\right) $
expresses precisely the exact invariance of the Gribov-Zwanziger action $S_{%
\mathrm{GZ}}$ under the operator $s_{\gamma }$
\begin{equation}
s_{\gamma }S_{\mathrm{GZ}}=0\ ,  \label{sc2-18}
\end{equation}%
where $s_{\gamma }$ is given by
\begin{eqnarray}
s_{\gamma }A_{\mu }^{a} &=&-\left( D_{\mu }c\right) ^{a}\ ,  \notag \\
s_{\gamma }c^{a} &=&\frac{1}{2}gf^{abc}c^{b}c^{c}\;,  \notag \\
s_{\gamma }\overline{c}^{a}(x) &=&b^{a}(x)+\int d^{4}y\ \left( D_{\nu
}^{mc}\Lambda _{\nu }^{c}\right) _{y}\left( \mathcal{M}^{-1}\right)
_{yx}^{ma}\ ,  \notag \\
s_{\gamma }b^{a} &=&0\;,  \notag \\
s_{\gamma }\overline{\omega }_{\mu }^{ab}(x) &=&\overline{\varphi }_{\mu
}^{ab}(x)+\gamma ^{2}g\int d^{4}y\ f^{mpb}A_{\mu }^{m}(y)\left( \mathcal{M}%
^{-1}\right) _{yx}^{pa}\;,  \notag \\
s_{\gamma }\overline{\varphi }_{\mu }^{ab} &=&0\ ,  \notag \\
s_{\gamma }\varphi _{\mu }^{ab} &=&\omega _{\mu }^{ab}\ ,  \notag \\
s_{\gamma }\omega _{\mu }^{ab} &=&0\ .  \label{sc2-19}
\end{eqnarray}%
Like the operator $s_{\vartheta }$ of eq.$\left( \ref{a20}\right)
$, the modified operator $s_{\gamma }$ is non-local. However,
unlike $s_{\vartheta }$, it lacks nilpotency. Nevertheless, it
turns out to be helpful in order to evaluate the vacuum
expectation value of BRST\ exact quantities, like the condensate
$\left\langle 0\left\vert \left( \overline{\varphi }_{\mu
}^{ab}(x)\varphi _{\mu }^{ab}(x)-\overline{\omega }_{\mu
}^{ab}(x)\omega _{\mu }^{ab}(x)\right) \right\vert 0\right\rangle
$.

\subsection{Evaluation of $\left\langle 0\left\vert \left( \overline{\protect%
\varphi }_{\protect\mu }^{ab}(x)\protect\varphi _{\protect\mu }^{ab}(x)-\overline{%
\protect\omega }_{\protect\mu }^{ab}(x)\protect\omega _{\protect\mu %
}^{ab}(x)\right) \right\vert 0\right\rangle $}

To evaluate the condensate $\left\langle 0\left\vert \left( \overline{%
\varphi }_{\mu }^{ab}(x)\varphi _{\mu }^{ab}(x)-\overline{\omega
}_{\mu }^{ab}(x)\omega _{\mu }^{ab}(x)\right) \right\vert
0\right\rangle $ we start from the condition
\begin{equation}
\left\langle 0\left\vert s_{\gamma }\left( \overline{\omega }_{\mu
}^{ab}(x)\varphi _{\mu }^{ab}(x)\right) \right\vert 0\right\rangle \ =0\ .
\label{sc2-20}
\end{equation}%
Making use of transformations $\left( \ref{sc2-19}\right) $, we get
\begin{equation}
\left\langle 0\left\vert s_{\gamma }\left( \overline{\omega }_{\mu
}^{ab}(x)\varphi _{\mu }^{ab}(x)\right) \right\vert 0\right\rangle \
=\left\langle 0\left\vert \left( \overline{\varphi }_{\mu }^{ab}\varphi
_{\mu }^{ab}-\overline{\omega }_{\mu }^{ab}\omega _{\mu }^{ab}\right)
\right\vert 0\right\rangle \ +\gamma ^{2}gf^{abc}\int d^{4}y\ \left\langle
0\left\vert A_{\mu }^{a}(y)\left( \mathcal{M}^{-1}\right) _{yx}^{bm}\varphi
_{\mu }^{mc}(x)\right\vert 0\right\rangle  \label{sc2-21}
\end{equation}
Therefore
\begin{equation}
\left\langle 0\left\vert \left( \overline{\varphi }_{\mu }^{ab}(x)\varphi _{\mu
}^{ab}(x)-\overline{\omega }_{\mu }^{ab}(x)\omega _{\mu }^{ab}(x)\right) \right\vert
0\right\rangle \ =-\gamma ^{2}gf^{abc}\int d^{4}y\ \left\langle 0\left\vert
A_{\mu }^{a}(y)\left( \mathcal{M}^{-1}\right) _{yx}^{bm}\varphi _{\mu
}^{mc}(x)\right\vert 0\right\rangle \ .  \label{sc2-22}
\end{equation}
Equation $\left( \ref{sc2-22}\right)$ is a non-perturbative Ward
identity relating the dimension two condensate to the Green's
function appearing in the right-hand side. One should also notice
that expression $\left( \ref{sc2-22}\right)$ displays explicit
dependence from the Gribov parameter $\gamma$,
as expected from the non-perturbative nature of $\left\langle 0\left\vert \left( \overline{%
\varphi }_{\mu }^{ab}\varphi _{\mu }^{ab}-\overline{\omega }_{\mu
}^{ab}\omega _{\mu }^{ab}\right) \right\vert 0\right\rangle $.
\\\\To the first order approximation
\begin{equation}
\left( \mathcal{M}^{-1}\right) _{yx}^{bm}=\left( \frac{1}{-\partial ^{2}}%
\right) _{yx}\delta ^{bm}+...\ \ ,  \label{sc2-23}
\end{equation}%
so that
\begin{eqnarray}
\left\langle 0\left\vert \left( \overline{\varphi }_{\mu }^{ab}\varphi _{\mu
}^{ab}-\overline{\omega }_{\mu }^{ab}\omega _{\mu }^{ab}\right) \right\vert
0\right\rangle &=&-\gamma ^{2}gf^{abc}\int d^{4}y\ \left\langle 0\left\vert
A_{\mu }^{a}(y)\left( \frac{1}{-\partial ^{2}}\right) _{yx}\varphi _{\mu
}^{bc}(x)\right\vert 0\right\rangle  \notag \\
&=&-\gamma ^{2}gf^{abc}\int d^{4}y\ \int \frac{d^{4}p}{\left( 2\pi \right)
^{4}}\frac{1}{p^{2}}e^{-ip(x-y)}\left\langle 0\left\vert A_{\mu
}^{a}(y)\varphi _{\mu }^{bc}(x)\right\vert 0\right\rangle  \notag \\
&=&-\gamma ^{2}gf^{abc}\int \frac{d^{4}p}{\left( 2\pi \right) ^{4}}\frac{1}{%
p^{2}}\left\langle 0\left\vert A_{\mu }^{a}(p)\varphi _{\mu
}^{bc}(-p)\right\vert 0\right\rangle \ .  \label{sc2-24}
\end{eqnarray}%
For the propagator $\left\langle 0\left\vert A_{\mu
}^{a}(p)\varphi _{\nu }^{bc}(-p)\right\vert 0\right\rangle $, we have \cite%
{Dudal:2008sp}
\begin{eqnarray}
\left\langle 0\left\vert A_{\mu }^{a}(p)\varphi _{\nu }^{bc}(-p)\right\vert
0\right\rangle &=&-\frac{f^{abc}\widehat{\gamma }^{2}}{\sqrt{2}\left( p^{4}+N%
\widehat{\gamma }^{4}\right) }\left( \delta _{\mu \nu }-\frac{p_{\mu }p_{\nu
}}{p^{2}}\right) \ ,  \label{sc2-25} \\
\widehat{\gamma }^{2} &=&\sqrt{2}g\gamma ^{2}\ .  \notag
\end{eqnarray}
Therefore
\begin{eqnarray}
\left\langle 0\left\vert \left( \overline{\varphi }_{\mu }^{ab}\varphi _{\mu
}^{ab}-\overline{\omega }_{\mu }^{ab}\omega _{\mu }^{ab}\right) \right\vert
0\right\rangle &=&\frac{3}{\sqrt{2}}\gamma ^{2}gf^{abc}f^{abc}\widehat{%
\gamma }^{2}\int \frac{d^{4}p}{\left( 2\pi \right) ^{4}}\frac{1}{p^{2}}\frac{%
1}{\left( p^{4}+N\widehat{\gamma }^{4}\right) }  \label{sc2-26} \\
&=&\frac{3\sqrt{2}}{64\pi }g\sqrt{N}(N^{2}-1)\gamma ^{2}\ ,  \notag
\end{eqnarray}
where use has been made of
\begin{equation}
f^{abc}f^{dbc}=N\delta ^{ad}\ .  \label{sc2-27}
\end{equation}
Expression $\left( \ref{sc2-26}\right) $ agrees with the result already
found in \cite{Dudal:2008sp}.

\subsection{Remarks on the condition $\left\langle 0\left\vert s_{\protect%
\gamma }\Theta(x) \right\vert 0\right\rangle =0\ $}

In this section we aim at providing a better understanding of the
condition $\left\langle 0\left\vert s_{\gamma }\Theta(x) \right\vert
0\right\rangle =0$. To that purpose, we shall study the
quantity $\left\langle 0\left\vert s_{\gamma }\left( gf^{abc}A_{\mu }^{a}(x)
\overline{\omega }_{\mu }^{bc}(x)\right) \right\vert 0\right\rangle $. \newline
\newline
From eqs.$\left( \ref{sc2-19}\right) $, we have
\begin{eqnarray}
\left\langle 0\left\vert s_{\gamma }\left( gf^{abc}A_{\mu }^{a}\overline{%
\omega }_{\mu }^{bc}\right) \right\vert 0\right\rangle &=&-\left\langle
0\left\vert \left( gf^{abc}\left( D_{\mu }^{ap}c^{p}\right) \overline{\omega
}_{\mu }^{bc}\right) \right\vert 0\right\rangle +\left\langle 0\left\vert
gf^{abc}A_{\mu }^{a}\overline{\varphi }_{\mu }^{bc}\right\vert 0\right\rangle
\notag \\
&&+\gamma ^{2}g^{2}f^{abc}\int d^{4}y\ f^{mpc}\left\langle 0\left\vert
A_{\mu }^{a}(x)A_{\mu }^{m}(y)\left( \mathcal{M}^{-1}\right)
_{yx}^{pb}\right\vert 0\right\rangle \ .  \label{sc2-29}
\end{eqnarray}
Taking into account that expression $\left\langle 0\left\vert \left( gf^{abc}\left(
D_{\mu }^{ap}c^{p}\right) \overline{\omega }_{\mu }^{bc}\right) \right\vert
0\right\rangle $ vanishes, due to the discrete symmetry $\left( \overline{%
\omega }_{\mu }^{bc}\rightarrow -\overline{\omega }_{\mu }^{bc}\ ,\ \omega
_{\mu }^{bc}\rightarrow -\omega _{\mu }^{bc}\right) $ of the action $\left( %
\ref{sc2-10}\right) $, one gets
\begin{equation}
\left\langle 0\left\vert s_{\gamma }\left( gf^{abc}A_{\mu }^{a}\overline{%
\omega }_{\mu }^{bc}\right) \right\vert 0\right\rangle =\left\langle
0\left\vert gf^{abc}A_{\mu }^{a}\overline{\varphi }_{\mu }^{bc}\right\vert
0\right\rangle +\gamma ^{2}g^{2}f^{abc}\int d^{4}y\ f^{mpc}\left\langle
0\left\vert A_{\mu }^{a}(x)A_{\mu }^{m}(y)\left( \mathcal{M}^{-1}\right)
_{yx}^{pb}\right\vert 0\right\rangle \ .  \label{sc2-30}
\end{equation}%
Reminding now the expression of the horizon function $h(x)$ of eq.$\left( %
\ref{sc2-4}\right) ,$ it follows $\;$
\begin{equation}
\left\langle 0\left\vert s_{\gamma }\left( gf^{abc}A_{\mu }^{a}\overline{%
\omega }_{\mu }^{bc}\right) \right\vert 0\right\rangle =\left\langle
0\left\vert gf^{abc}A_{\mu }^{a}\overline{\varphi }_{\mu }^{bc}\right\vert
0\right\rangle +\gamma ^{2}\left\langle 0\left\vert h(x)\right\vert
0\right\rangle \ =0\ ,  \label{inc2}
\end{equation}%
due to equation $\left( \ref{inc1}\right) $. \newline
\newline
We see thus that the equation $\left\langle 0\left\vert s_{\gamma }\left(
gf^{abc}A_{\mu }^{a}\overline{\omega }_{\mu }^{bc}\right) \right\vert
0\right\rangle =0$ follows from the gap equation $\left( \ref{inc1}\right) $
for the Gribov parameter $\gamma $. This provides support to the condition $%
\left\langle 0\left\vert s_{\gamma }\Theta(x) \right\vert 0\right\rangle =0$,
which seems to emerge in a natural way within the Gribov-Zwanziger framework.

\subsection{A general formula}

We can now establish a general procedure to evaluate the BRST\ exact
quantity $\left\langle 0\left\vert s\Theta(x) \right\vert 0\right\rangle $. Let
us decompose the operator $s_{\gamma }$ as
\begin{equation}
s_{\gamma }=s+\delta _{\gamma }\ ,  \label{g1}
\end{equation}%
where $s$ is the BRST\ operator of eqs.$\left( \ref{ii1}\right) $ and $%
\delta _{\gamma }$ stands for the operator
\begin{eqnarray}
\delta _{\gamma }A_{\mu }^{a} &=&0\ ,  \notag \\
\delta _{\gamma }c^{a} &=&0\;,  \notag \\
\delta _{\gamma }\overline{c}^{a}(x) &=&\int d^{4}y\ \left( D_{\nu
}^{mc}\Lambda _{\nu }^{c}\right) _{y}\left( \mathcal{M}^{-1}\right)
_{yx}^{ma}\ ,  \notag \\
\delta _{\gamma }b^{a} &=&0\;,  \notag \\
\delta _{\gamma }\overline{\omega }_{\mu }^{ab}(x) &=&\gamma ^{2}g\int
d^{4}y\ f^{mpb}A_{\mu }^{m}(y)\left( \mathcal{M}^{-1}\right) _{yx}^{pa}\;,
\notag \\
\delta _{\gamma }\overline{\varphi }_{\mu }^{ab} &=&0\ ,  \notag \\
\delta _{\gamma }\varphi _{\mu }^{ab} &=&0\ ,  \notag \\
\delta _{\gamma }\omega _{\mu }^{ab} &=&0\ .  \label{g2}
\end{eqnarray}
Therefore, from the condition
\begin{equation}
\left\langle 0\left\vert s_{\gamma }\Theta(x) \right\vert 0\right\rangle =0\ ,
\label{g3}
\end{equation}%
we get
\begin{equation}
\left\langle 0\left\vert s\Theta(x) \right\vert 0\right\rangle =-\left\langle
0\left\vert \delta _{\gamma }\Theta(x) \right\vert 0\right\rangle \ .
\label{g4}
\end{equation}%
We see thus that the characterization of quantities which are BRST\ exact
relies on the operator $\delta _{\gamma }$ which takes into account the
presence of the Gribov horizon through its explicit dependence from the
parameter $\gamma $.

\subsubsection{From condensates to correlation functions}

The formula $\left( \ref{g4}\right) $ enables us to extend the
previous construction to the case of correlation functions which
are BRST exact. As an example, let us evaluate the two-point
correlation function
\begin{equation}
\left\langle  \left( \overline{\varphi }_{\mu }^{ab}(x)\varphi _{\mu
}^{ab}(y)-\overline{\omega }_{\mu }^{ab}(x)\omega _{\mu }^{ab}(y)\right)
\right\rangle \ = \left\langle s
\left( \overline{\omega }_{\mu }^{ab}(x) \varphi _{\mu
}^{ab}(y) \right) \right\rangle \ . \label{n1}
\end{equation}
From the condition
\begin{equation}
\left\langle s_{\gamma} \left(
\overline{\omega }_{\mu }^{ab}(x) \varphi _{\mu }^{ab}(y) \right)
\right\rangle =0 \ , \label{n2}
\end{equation}
we easily get
\begin{equation}
\left\langle 0\left\vert \left( \overline{\varphi }_{\mu }^{ab}(x)\varphi _{\mu
}^{ab}(y)-\overline{\omega }_{\mu }^{ab}(x)\omega _{\mu }^{ab}(y)\right) \right\vert
0\right\rangle \ =-\gamma ^{2}gf^{mpb}\int d^{4}z\ \left\langle 0\left\vert
A_{\mu }^{m}(z)\left( \mathcal{M}^{-1}\right) _{zx}^{pa}\varphi _{\mu
}^{mc}(y)\right\vert 0\right\rangle \ .  \label{n2}
\end{equation}

\section{The refined Gribov-Zwanziger action}

The so called refined Gribov-Zwanziger action was introduced in \cite%
{Dudal:2007cw,Dudal:2008sp}. It takes into account the existence of the
dimension two condensate $\left\langle 0\left\vert \left( \overline{\varphi }%
_{\mu }^{ab}\varphi _{\mu }^{ab}-\overline{\omega }_{\mu }^{ab}\omega _{\mu
}^{ab}\right) \right\vert 0\right\rangle $, which is non-vanishing for
non-vanishing Gribov parameter $\gamma $. This condensate
reflects the non-trivial dynamics developed by the auxiliary fields $\left(
\overline{\varphi }_{\mu }^{ab},\varphi _{\mu }^{ab},\overline{\omega }_{\mu
}^{ab},\omega _{\mu }^{ab}\right) $. This feature has motivated the introduction of
the refined Gribov-Zwanziger action, in which the operator $\left( \overline{%
\varphi }_{\mu }^{ab}\varphi _{\mu }^{ab}-\overline{\omega }_{\mu
}^{ab}\omega _{\mu }^{ab}\right) $ is taken into account from the beginning.
This is achieved by adding to the Gribov-Zwanziger action the term
\begin{equation}
S_{\overline{\varphi }\varphi }=-\int d^{4}x\ \mu ^{2}\left( \overline{%
\varphi }_{\mu }^{ab}\varphi _{\mu }^{ab}-\overline{\omega }_{\mu
}^{ab}\omega _{\mu }^{ab}\right) \ ,  \label{sc3-1}
\end{equation}%
which is left invariant by the BRST\ transformations $\left( \ref{ii1}%
\right) $. The parameter $\mu ^{2}$ is a mass parameter which, as much as
the Gribov parameter $\gamma $, is fixed by a variational principle, see
\cite{Dudal:2007cw,Dudal:2008sp}\footnote{The parameter $\mu$ is
determined by requiring that the Gribov no-pole condition \cite{Gribov:1977wm} is
fulfilled in the presence of the new operator $\left( \ref{sc3-1} \right)$, so that
the Gribov horizon is not crossed. In other words, the introduction
of the term $\left( \ref{sc3-1} \right) $ can be done in such a way that one
remains within the Gribov region \cite{Dudal:2007cw,Dudal:2008sp}}. The refined Gribov-Zwanziger action
is thus defined by
\begin{equation}
S_{\mathrm{RGZ}}=S_{\mathrm{GZ}}+S_{\overline{\varphi }\varphi }\ .
\label{sc3-2}
\end{equation}%
Since the term $S_{\overline{\varphi }\varphi }$ is BRST\ invariant, the
action $S_{\mathrm{RGZ}}$ displays the same breaking term of the Gribov-Zwanziger
action $S_{\mathrm{GZ}}$, \textit{i.e. }
\begin{eqnarray}
sS_{\mathrm{RGZ}} &=&\int d^{4}x\ gf^{abm}\left( D_{\nu }^{bp}c^{p}\right)
\left( \left( \partial _{\nu }\overline{\varphi }_{\mu }^{ac}\right) \varphi
_{\mu }^{mc}-\left( \partial _{\nu }\overline{\omega }_{\mu }^{ac}\right)
\omega _{\mu }^{mc}\right)   \notag \\
&&-\gamma ^{2}g\int d^{4}x\ \left( c^{m}D_{\mu }^{ma}\left( f^{abc}\left(
\varphi _{\mu }^{bc}+\overline{\varphi }_{\mu }^{bc}\right) \right)
+f^{abc}A_{\mu }^{a}\omega _{\mu }^{bc}\right) \ .  \label{sc3-3}
\end{eqnarray}%
The refined Gribov-Zwanziger action turns out to be renormalizable to all
orders, due to the large set of Ward identities which can be established \cite%
{Dudal:2007cw,Dudal:2008sp}. It is worth mentioning that the action $\left( %
\ref{sc3-2}\right) $ gives rise to a decoupling type solution for the ghost and
gluon propagator. The ghost propagator turns out to behave as $1/k^2$ in the infrared,
while the gluon propagator does not vanish at the
origin in momentum space, exhibiting positivity violation\footnote{%
The positivity violation of the gluon propagator invalidates the
interpretation of the gluons as excitations of the physical spectrum of the
theory, providing a simple understanding of gluon confinement.}, namely
\begin{equation}
\left\langle A_{\mu }^{a}(p)A_{\nu }^{b}(-p)\right\rangle =\delta ^{ab}\frac{%
p^{2}+\mu ^{2}}{p^{4}+\mu ^{2}p^{2}+2g^{2}N\gamma ^{4}}\left( \delta _{\mu
\nu }-\frac{p_{\mu }p_{\nu }}{p^{2}}\right) \ .  \label{sc3-33}
\end{equation}%
The infrared behavior displayed by expression $\left(
\ref{sc3-33}\right) $ is in good agreement with the most recent
numerical simulations performed on huge lattices
\cite{Cucchieri:2007rg,Cucchieri:2008fc,Bornyakov:2008yx,Bogolubsky:2009dc,Cucchieri:2009zt},
as well as with the
analytical results found in \cite{Aguilar:2008xm,Boucaud:2008ky}.
Let us underline that, without the introduction of the operator
$\left( \ref{sc3-1}\right) $, the Gribov-Zwanziger action would
give rise to a different behavior for the propagators, namely to
an enhanced ghost and to a gluon propagator which would vanish at
the origin, see also the recent discussion presented in
\cite{Zwanziger:2009je}.
\newline
\newline
Let us proceed by showing how the operator $s_{\gamma }$
generalizes to the present case. From the refined Gribov-Zwanziger
action, we get
\begin{eqnarray}
\frac{\delta S_{\mathrm{RGZ}}}{\delta \overline{c}^{a}} &=&-\mathcal{M}%
^{ab}c^{b}\ ,  \notag \\
\frac{\delta S_{\mathrm{RGZ}}}{\delta \overline{\omega }_{\mu }^{ac}}
&=&\left( \mathcal{M}^{ab}+\mu ^{2}\delta^{ab}\right) \omega _{\mu }^{bc}\mathcal{\ }%
\mathrm{,}  \label{sc3-4}
\end{eqnarray}%
so that
\begin{eqnarray}
c^{b}(x) &=&-\left( \mathcal{M}^{-1}\right) _{xy}^{bc}\frac{\delta S_{%
\mathrm{RGZ}}}{\delta \overline{c}_{y}^{c}}=-\int d^{4}y\ \left[ \left(
\mathcal{M}^{-1}(x,y)\right) ^{bc}\right] \frac{\delta S_{\mathrm{RGZ}}}{%
\delta \overline{c}^{c}(y)}\ ,  \notag \\
\omega _{\mu }^{dc}(x) &=&\left( \left( \mathcal{M+}\mu ^{2}\right)
^{-1}\right) _{xy}^{da}\frac{\delta S_{\mathrm{RGZ}}}{\delta \overline{%
\omega }_{y\mu }^{ac}}=\int d^{4}y\ \left[ \left( \left( \mathcal{M}%
\mathcal{+}\mu ^{2}\right)^{-1}\right)_{xy} ^{da}\right] \frac{\delta S_{\mathrm{%
RGZ}}}{\delta \overline{\omega }^{ac}(y)}\ .  \label{sc3-5}
\end{eqnarray}%
Thus, equation $\left( \ref{sc3-3}\right) $ can be rewritten as
\begin{equation}
sS_{\mathrm{RGZ}}=\int d^{4}x\left( -\left( D_{\nu }^{ma}\Lambda _{\nu
}^{a}\right) _{x}\left( \mathcal{M}^{-1}\right) _{xy}^{mc}\frac{\delta S_{%
\mathrm{RGZ}}}{\delta \overline{c}_{y}^{c}}-\gamma ^{2}gf^{abc}A_{\mu
}^{a}(x)\left( \left( \mathcal{M+}\mu ^{2}\right) ^{-1}\right) _{xy}^{bm}%
\frac{\delta S_{\mathrm{RGZ}}}{\delta \overline{\omega }_{y\mu }^{mc}}%
\right) \ ,  \label{sc3-6}
\end{equation}%
with $\Lambda _{\nu }^{a}$ given in expression $\left( \ref{sc2-17}\right) $%
. Eq.$\left( \ref{sc3-6}\right) $ expresses the exact invariance of the
refined Gribov-Zwanziger action $S_{\mathrm{RGZ}}$ under the nonlocal
operator $s_{\gamma}^{(\mu)}$
\begin{equation}
s_{\gamma}^{(\mu)} S_{\mathrm{RGZ}}=0\ ,  \label{sc3-8}
\end{equation}%
where the operator $s_{\gamma}^{(\mu)}$ is now given by
\begin{eqnarray}
s_{\gamma}^{(\mu)} A_{\nu }^{a} &=&-\left( D_{\nu }c\right) ^{a}\ ,  \notag \\
s_{\gamma}^{(\mu)} c^{a} &=&\frac{1}{2}gf^{abc}c^{b}c^{c}\;,  \notag \\
s_{\gamma}^{(\mu)} \overline{c}^{a}(x) &=&b^{a}(x)+\int d^{4}y\ \left( D_{\nu
}^{mc}\Lambda _{\nu }^{c}\right) _{y}\left( \mathcal{M}^{-1}\right)
_{yx}^{ma}\ ,  \notag \\
s_{\gamma}^{(\mu)} b^{a} &=&0\;,  \notag \\
s_{\gamma}^{(\mu)} \overline{\omega }_{\nu }^{ab}(x) &=&\overline{\varphi }_{\nu
}^{ab}(x)+\gamma ^{2}g\int d^{4}y\ f^{mpb}A_{\nu }^{m}(y)\left( \left(
\mathcal{M+}\mu ^{2}\right) ^{-1}\right) _{yx}^{pa}\;,  \notag \\
s_{\gamma}^{(\mu)} \overline{\varphi }_{\nu }^{ab} &=&0\ ,  \notag \\
s_{\gamma}^{(\mu)} \varphi _{\nu }^{ab} &=&\omega _{\nu }^{ab}\ ,  \notag \\
s_{\gamma}^{(\mu)} \omega _{\nu }^{ab} &=&0\ .  \label{sc3-9}
\end{eqnarray}
The same calculations performed in the previous sections with the
operator $s_{\gamma}$ within the Gribov-Zwanziger action, can be
repeated by employing the operator $s_{\gamma}^{(\mu)}$ and by
taking as starting action the refined Gribov-Zwanziger action.

\section{Conclusion}

$\bullet$ In this work we have addressed the issue of the BRST\
symmetry in presence of the Gribov horizon. Our main results
are expressed by equations $\left( \ref{sc2-18}\right)-\left( \ref{sc2-19}%
\right) $, which show that, due to the positivity of the
Faddeev-Popov operator ${\mathcal M}^{ab}$ within the Gribov
region $\Omega$, the soft breaking of the BRST\ symmetry exhibited
by the Gribov-Zwanziger action can be converted into an exact
invariance. The resulting modified operator $s_{\gamma }$ displays
an explicit dependence from the Gribov parameter $\gamma $. We
have shown, through few examples, that the operator $s_{\gamma }$
might be helpful in order to evaluate the vacuum expectation value
of quantities which are BRST\ exact. \\\\As far as the
renormalizability properties of the Gribov-Zwanziger action are
concerned, the operator $s_\gamma$ has not to be considered as a
substitute of the BRST operator s. Albeit broken, we remind here
that the BRST transformations $\left( \ref{ii1} \right)$ give rise
to a set of softly broken Slavnov-Taylor identities which are
perfectly suitable for a study of the renormalizability of the
theory within a local framework
\cite{Zwanziger:1989mf,Zwanziger:1992qr,Maggiore:1993wq,Dudal:2005na,Dudal:2008sp,Gracey:2006dr}.
Rather, the operator $s_\gamma$ has to be seen as a useful device
to extract non-perturbative Ward identities in order to
characterize the vacuum expectation value of quantities which are
BRST exact.
\\\\$\bullet$ It is worth spending here
a few words about the Gribov-Zwanziger (GZ) action and the
so-called refined Gribov-Zwanziger action (RGZ) introduced in
\cite{Dudal:2007cw,Dudal:2008sp}. Both GZ and RGZ models are
renormalizable and exhibit the same BRST soft breaking.  Moreover,
as it is apparent from the expression $\left( \ref{sc3-1} \right)
$, the GZ and RGZ models differ by a BRST exact term, notably
\begin{equation}
\left( \overline{\varphi }_{\mu }^{ab}\varphi _{\mu }^{ab}-\overline{\omega }_{\mu
}^{ab}\omega _{\mu }^{ab}\right) =  s \left( \overline{\omega }_{\mu }^{ab}\varphi _{\mu }^{ab}\right)
\ .  \label{f1}
\end{equation}
As already mentioned, the introduction of this operator in the RGZ action
is supported by the non-vanishing value of the condensate $\left\langle 0\left\vert \left( \overline{\varphi }%
_{\mu }^{ab}\varphi _{\mu }^{ab}-\overline{\omega }_{\mu
}^{ab}\omega _{\mu }^{ab}\right) \right\vert 0\right\rangle $,
whose non-perturbative effects are taken into account from the
beginning, through the introduction of expression $\left( \ref{f1}
\right)$. We point out that the operator $\left( \ref{f1} \right)
$ is the only operator which can be added to the GZ action without
jeopardizing the renormalizability, while keeping the same BRST
breaking as well as the same field content. In the absence of the
Gribov horizon, and thus in the case of an unbroken BRST symmetry,
the two actions would be physically equivalent. However, in the
present case in which the BRST symmetry is softly broken, they
give rise to rather different predictions, both in four and three
dimensions. In fact, while the GZ action predicts an infrared
enhanced ghost sector and a gluon propagator which vanishes at the
origin \cite{Zwanziger:2009je}, the output of the RGZ action is a
ghost propagator which is not enhanced in the infrared and a gluon
propagator which does not vanish at the origin in momentum space.
So far, the behavior predicted by the RGZ model seems to be in
good agreement with the recent numerical data
\cite{Cucchieri:2007rg,Cucchieri:2008fc,Bornyakov:2008yx,Bogolubsky:2009dc,Cucchieri:2009zt}.
Interestingly, in two dimensions, both models coalesce
\cite{Dudal:2008xd}\footnote{This is due to fact that in 2d the
operator $\left( \ref{f1} \right)$ cannot be safely introduced
\cite{Dudal:2008xd}, due to the appearance of infrared
divergencies in the Feynman integrals, typical of 2d theories. We
notice in fact that in 2d the auxiliary fields $(\overline{\varphi
}_{\mu }^{ab},\varphi _{\mu }^{ab},\overline{\omega }_{\mu
}^{ab},\omega _{\mu }^{ab})$ are dimensionless.}, giving an
enhanced ghost and a vanishing gluon propagator, in agreement with
the lattice data in two dimensions
\cite{Maas:2007uv,Cucchieri:2007rg,Cucchieri:2009zt}. At present,
we are unable to give a simple explanation of the different
behavior predicted by the GZ and RGZ actions in three and four
dimensions.  We believe that a better understanding of the meaning
of the BRST breaking and of its deep connection with the Gribov
horizon is needed.
\\\\$\bullet$ Another issue which deserves further investigation is that of the role played
by the auxiliary fields $\left( \overline{\varphi }_{\mu }^{ab},
\varphi _{\mu }^{ab}, \overline{\omega }_{\mu }^{ab}, \omega _{\mu
}^{ab}\right)$ within the Gribov-Zwanziger framework. As already noticed, 
the introduction of these fields stems from
the necessity of localizing the horizon function, eq.\eqref{m1},
so that a local and renormalizable action is obtained from the
restriction to the Gribov region $\Omega$. These
fields are thus expected to carry nontrivial information about the
Gribov horizon, a fact which can be already inferred from the
explicit dependence of the dimension two condensate $\langle
\overline{\varphi }_{\mu }^{ab} \varphi _{\mu }^{ab}-
\overline{\omega }_{\mu }^{ab} \omega _{\mu }^{ab}\rangle$ from
the Gribov parameter $\gamma$, eq.\eqref{i1}. Also, as the
restriction to the Gribov region deeply modifies the infrared
behavior of the theory, we believe that a better understanding of
the dynamics of the auxiliary fields could provide useful
information about the confining character of the theory as well as
about the construction of its physical spectrum. It is worth therefore 
to add here a few remarks on the on-going investigation \cite{wp} about
the role which these fields might play. For illustration purposes, let us
rely on an explicit example taken from the Gribov-Zwanziger
theory, see \cite{Zwanziger:1989mf}. Let us start by giving a look
at the gluon propagator of the Gribov-Zwanziger theory, which can
be obtained from the expression \eqref{sc3-33} by setting $\mu
=0$, namely
\begin{equation}
\left\langle A_{\mu }^{a}(k)A_{\nu }^{b}(-k)\right\rangle =\delta
^{ab}\left( \delta _{\mu \nu }-\frac{k_{\mu }k_{\nu }}{k^{2}}\right) \frac{%
k^{2}}{k^{4}+{\hat \gamma }^{4}}\;, \qquad {\hat{\gamma}}^4 = 2 g^2 N \gamma^4 \;,  \label{z1}
\end{equation}%
As it is apparent, expression \eqref{z1} displays complex poles at
$k^{2}=\pm i{\hat \gamma}^{2}$, which invalidates the
interpretation of gluons as excitations of the physical spectrum.
In other words, in the infrared, gluons cannot be considered as
part of the physical spectrum: they are confined. The physical
excitations of the theory would correspond to colorless bound
states of gluons, \textit{i.e.} glueballs. We expect thus that
information about the physical spectrum could be obtained by
looking at the correlation function of colorless gauge invariant
operators as, for example:
\begin{equation}
G(p)=\int d^{4}x\ e^{-ipx\ }\left\langle F^{2}(x)F^{2}(0)\right\rangle \ ,
\label{z2}
\end{equation}
where $F^2(x)=F^a_{\mu\nu}(x)F^a_{\mu\nu}(x)$. This correlation function is useful in order to
investigate the properties of the scalar spin zero glueball state. Within the Gribov-Zwanziger theory,
it has to be evaluated with the Feynman rules obtained by employing
the confining gluon propagator \eqref{z1}. The explicit first order evaluation of expression
\eqref{z2} can be found in \cite{Zwanziger:1989mf}, and can be summarized as follows:
\begin{equation}
G(p) = G^{\rm phys}(p) + G^{\rm unphys}(p) \;. \label{z3}
\end{equation}
The unphysical part, $G^{\rm unphys}(p)$, displays cuts beginning
at the unphysical values $p^2=\pm 4i{\hat \gamma}^2$, whereas the
physical part, $G^{\rm phys}(p)$, has a cut beginning at the
physical threshold $p^2=-2{\hat \gamma}^2$. Moreover, the spectral
function of $G^{\rm phys}(p)$ turns out to be positive, so that it
obeys the Kallen-Lehmann representation \cite{Zwanziger:1989mf}.
As such, $G^{\rm phys}(p)$ is an acceptable correlation function
for physical glueball excitations. What is interesting in
expression \eqref{z3} is that a physical cut has emerged in the
correlation function of gauge invariant quantities, even if it has
been evaluated with a gluon propagator exhibiting only unphysical
complex poles. This is precisely what one expects from a confining
theory. Gluons are not physical excitations, but the glueball
correlation function displays a physical singularity.
\\\\Of course, we have still to face the very difficult presence of the
unphysical part $G^{\rm unphys}(p)$. Here, we can argue that the
presence of the auxiliary fields $\left( \overline{\varphi }_{\mu
}^{ab}, \varphi _{\mu }^{ab}, \overline{\omega }_{\mu }^{ab},
\omega _{\mu }^{ab}\right)$ might be a welcome feature. In fact,
we can figure out that, in the presence of the Gribov horizon, the
correlation functions of the physical operators should receive
contributions from the horizon, which could be accounted for by
the auxiliary fields. This reasoning can be made on a quantitative
basis by recalling that in ordinary Yang-Mills theories the
construction of the physical operators is obtained through the
cohomology of the nilpotent BRST operator
\cite{Becchi:2005dr,Piguet:1995er}. It is a remarkable fact that,
in the present case, the BRST operator $s$ of eqs.\eqref{ii1}
remains nilpotent, so that physical operators can be still
associated to its cohomology classes. Furthermore, one should
observe that the auxiliary fields $\left( \overline{\varphi }_{\mu
}^{ab}, \varphi _{\mu }^{ab}, \overline{\omega }_{\mu }^{ab},
\omega _{\mu }^{ab}\right)$  are assembled in BRST doublets
\cite{Piguet:1995er}, so that they can only appear through terms
which are BRST exact. Therefore, for the glueball operator $O_{\it
glueb}(x)$ we would write
\begin{equation}
O_{\it glueb}(x) = F^{2}(x)  +  s{\cal R}_{\gamma}(x) \;,
\label{goper}
\end{equation}
where ${\cal R}_{\gamma}$ is the exact BRST piece which depends on
the auxiliary fields\footnote{See also the recent work \cite{Dudal:2009zh}.}.  
Expression \eqref{goper} is easily seen to be BRST
invariant, namely
\begin{equation}
s O_{\it glueb}(x) = 0 \;. \label{ex}
\end{equation}
For the glueball correlation function, we would get
\begin{equation}
\langle O_{\it glueb}(x) O_{\it glueb}(y) \rangle = \langle F^{2}(x) F^{2}(y) \rangle
 + \langle s \Lambda(x,y) \rangle \; \label{corr}
\end{equation}
where $\Lambda(x,y)$ stands for
\begin{equation}
\Lambda(x,y) = {\cal R}_{\gamma}(x) F^2(y) + F^2(x) {\cal R}_{\gamma}(y) + {\cal R}_{\gamma}(x) s({\cal R}_{\gamma}(y)) \;. \label{lambda}
\end{equation}
In the absence of the Gribov horizon, {\it i.e.} when $\gamma=0$,
the inclusion of the BRST exact term $ s{\cal R}_{\gamma}(x) $ is
completely irrelevant, due to the fact the BRST symmetry is
unbroken. As a consequence, $\langle s \Lambda(x,y) \rangle =0$,
so that the correlator $\langle O_{\it glueb}(x) O_{\it glueb}(y)
\rangle$ reduces to $\langle F^{2}(x) F^{2}(y)$. However, in the
presence of the Gribov horizon, {\it i.e.} for $\gamma \neq 0$,
BRST symmetry is softly broken, so that the quantity $\langle s
\Lambda(x,y) \rangle$ is now non-vanishing. We could thus search
for a suitable term  ${\cal R}_{\gamma}(x)$ which would enable us
to cancel, order by order, the unphysical contributions contained
in $G^{\rm unphys}(p)$. Thus, the correlation function $\langle
O_{\it glueb}(x) O_{\it glueb}(y) \rangle$ would display only
physical poles, corresponding to colorless gluon bound states. At
the present stage, such a possibility  has to be regarded as an
interesting plausible working hypothesis \cite{wp}. Certainly, it
is a point worth to be investigated. If true, this would mean that
the Gribov horizon combines in a nice and friendly fashion with
the soft breaking of the BRST operator in such a way that the
unphyiscal cuts present in the correlation functions of physical
operators could be removed order by order. The presence of the
auxiliary fields would be thus a quite welcome feature, as they
would give us a way to construct the exact BRST term ${\cal
R}_{\gamma}(x)$ needed to account for the unphysical cuts. At the
same time, the existence of the operator $s_\gamma$ would enable
us to evaluate the quantity $\langle s \Lambda(x,y) \rangle $ in a
simple and efficient way. \\\\$\bullet$ A topic which we also
intend to analyse is that of the possible existence of a
relationship between the non-local operator $s_\gamma$ and the
color invariance of the Gribov-Zwanziger action. This could open a
small window towards a better understanding of the color symmetry
in the presence of the Gribov horizon.\\\\$\bullet$ Finally, it is
worth mentioning that the construction of the non-local operator
$s_{\gamma}$ outlined here can be generalized to the maximal
Abelian gauge
\cite{'tHooft:1981ht,Kronfeld:1987vd,Min:1985bx,Fazio:2001rm,Dudal:2004rx,Capri:2008ak},
for which a study of the properties of the corresponding Gribov
region is available \cite{Capri:2008vk}.

\section*{Acknowledgments}
It is a pleasure to thank  Laurent Baulieu, Attilio Cucchieri, David Dudal, Nele
Vandersickel and Daniel Zwanziger for many interesting
discussions.  The Conselho Nacional de Desenvolvimento
Cient\'{\i}fico e Tecnol\'{o}gico (CNPq-Brazil), the Faperj,
Funda{\c{c}}{\~{a}}o de Amparo {\`{a}} Pesquisa do Estado do Rio
de Janeiro, the SR2-UERJ and the Coordena{\c{c}}{\~{a}}o de
Aperfei{\c{c}}oamento de Pessoal de N{\'{\i}}vel Superior (CAPES),
the CLAF, Centro Latino-Americano de F{\'\i}sica, are gratefully
acknowledged for financial support.

\end{document}